\documentclass[letter]{aa}

\usepackage{graphicx}
\usepackage{txfonts}
\usepackage{hyperref}
\usepackage{url}
\usepackage{xcolor}
\hypersetup{
colorlinks,
linkcolor={red!80!black},
citecolor={blue!80!black},
urlcolor={blue!80!black}
}

\def\cm3{cm$^{-3}$}
\def\kms{km~s$^{-1}$}

\def\rsun{R$_{\odot}$}

\def\msun{M$_{\odot}$}

\def\one{\ts {\,\sc i}}
\def\two{\ts {\,\sc ii}}

\def\beq{\begin{equation}}
\def\eeq{\end{equation}}

\def\lesssim{\mathrel{\hbox{\rlap{\hbox{\lower4pt\hbox{$\sim$}}}\hbox{$<$}}}}
\def\gtrsim{\mathrel{\hbox{\rlap{\hbox{\lower4pt\hbox{$\sim$}}}\hbox{$>$}}}}

\def\one{{\,\sc i}}
\def\two{{\,\sc ii}}

\def\v1d{{\sc v1d}}

\def\mesa{{\sc mesa}}
\def\cmfgen{{\sc cmfgen}}

\newcommand{\iso}[2]{\ensuremath{^{#1}\rm{#2}}}

\def\apj{ApJ}
\def\apjs{ApJS}
\def\apjl{ApJL}
\def\aap{A\&A}

\def\mnras{MNRAS}
\def\nat{Nature}

\def\nifs{\iso{56}Ni}

\begin{document}

   \title{A magnetar model for the hydrogen-rich super-luminous supernova iPTF14hls}
   \titlerunning{A magnetar model for iPTF14hls}

\author{Luc Dessart}

\institute{
Unidad Mixta Internacional Franco-Chilena de Astronom\'ia (CNRS, UMI 3386),
    Departamento de Astronom\'ia, Universidad de Chile,
    Camino El Observatorio 1515, Las Condes, Santiago, Chile
  }

   \date{Received; accepted}

  \abstract{
Transient surveys have recently revealed the existence of
H-rich super-luminous supernovae (SLSN; e.g., iPTF14hls, OGLE-SN14-073)
characterized by an exceptionally large time-integrated bolometric luminosity,
a sustained blue optical color, and Doppler-broadened H\one\ lines
at all times. Here, I investigate the effect that a magnetar
(initial rotational energy of $4 \times 10^{50}$\,erg and field strength of
$7 \times 10^{13}$\,G)
would have on the properties of a typical Type II SN ejecta (mass of 13.35\,\msun,
kinetic energy of $1.32 \times 10^{51}$\,erg, 0.077\,\msun\ of \nifs) produced
by the terminal explosion of an H-rich blue-supergiant star.
I present a non-LTE time-dependent radiative transfer simulation of
the resulting photometric and spectroscopic evolution from 1\,d until 600\,d
after explosion.
With magnetar power,  the model luminosity and brightness are enhanced, the ejecta
is everywhere hotter and more ionised, and the spectrum formation region is
much more extended.
This magnetar-powered SN ejecta reproduces
most of the observed properties of SLSN iPTF14hls, including the sustained
brightness of $-18$\,mag in the $R$ band, the blue optical color, and the broad
H\one\ lines for 600\,d.
The non-extreme magnetar properties, combined with the standard Type II SN
ejecta properties offer an interesting alternative to the pair-unstable super-massive
star model recently proposed, which involves a highly-energetic and super-massive ejecta.
Hence, such Type II SLSNe  may differ from standard Type
II SNe exclusively through the influence of a magnetar.
  }

   \maketitle

\section{Introduction}

    Super-luminous supernovae (SLSNe) owe their exceptional instantaneous
and/or time-integrated luminosities to a non-standard source of energy and power.
This power source may be interaction between a (standard-energy) ejecta with dense,
massive, and slow-moving circumstellar material, leading to an interacting
SN, generally of Type IIn (H-rich; \citealt{schlegel_90}; \citealt{chugai_98S_01};
\citealt{smith_06gy_07}; \citealt{moriya_rsg_csm_11};  \citealt{fransson_10jl};
\citealt{D15_2n}; \citealt{chugai_15}; \citealt{D16_2n}).
The spectral signatures are unambiguous, with
the presence of electron-scattering, rather than Doppler, broadened emission lines.
Alternatively, this power source may be a greater than standard production of
unstable isotopes, and in particular \nifs, as in pair-instability SNe
from super-massive stars \citep{barkat_pisn_67}. The large metal content of
these ejecta produce strongly blanketed, red, spectra with small/moderate
line widths at and beyond maximum \citep{d13_pisn}.
The final alternative is energy injection from a compact remnant,
as in a strongly magnetized neutron star with a fast initial spin.
For moderate magnetic field strengths and initial spin periods,
the spin-down time scale
may be equal to or greater than the expansion time scale of the ejecta,
allowing a powerful heating on day/week time scales \citep{KB10}.
This engine is believed to be at the origin of most, and perhaps all, SLSN Ic,
characterized by relatively short rise times, blue colors at all times, and the dominance
of spectral lines from intermediate mass elements like oxygen
\citep{quimby_slsnic_11,d12_magnetar,nicholl_slsn_13,greiner_pm_15,mazzali_slsn_16,
chen_slsnic_2d_17}.
Because of the nature of these processes, SLSNe should generally be
connected to core-collapse SNe.

\citet[hereafter A17]{arcavi_iptf14hls} recently reported the unique properties
of the Type II SLSN iPTF14hls.
This event has an inferred $R$ band
absolute magnitude of $-18$\,mag for about 600\,d
(inferred time-integrated bolometric luminosity of $2.2 \times 10^{50}$\,erg),
with fluctuations of amplitude 0.5\,mag. Its color is
blue throughout these two years, with $V-I \sim 0.2$\,mag.
The optical spectra of SLSN iPTF14hls evolve little
from about 100 to 600\,d after
the inferred (but uncertain) time of explosion. H$\alpha$, which
is the strongest line in the spectrum, evolves little in strength (relative
to the adjacent continuum) and in width.
A17 infer an H$\alpha$ formation region that
is much more extended that the radius of continuum formation, and propose
that this external region corresponds to a massive shell ejected a few hundred
days before a terminal explosion. In this context, iPTF14hls would be associated
with a pair-unstable super-massive star.

In this configuration, the inner ejecta from a terminal explosion would ram
into a massive (e.g., 50\,\msun) energetic (e.g., 10$^{52}$\,erg) outer shell with
a mean mass-weighted velocity of $\sim\,4000$\,\kms\ and located
at $10^{15} - 10^{16}$\,cm. Electron-scattering broadened narrow
lines do not form since photons from the shock are reprocessed in a fast
outer shell in homologous expansion. This model is the high-energy
counterpart to the proposed model for SN\,1994W analogs
\citep{chugai_15,D16_2n}. The interaction leads to the formation
of a heat wave that propagates outwards in the outer shell, causing
reionization, and shifting the photosphere to large radii (or velocities).
After a bolometric
maximum reached on a diffusion time scale, the outer shell recombines
and the photosphere recedes in mass/velocity space.
Compared to interaction with a slow long-lived dense
wind, the interaction with a massive energetic explosively ejected outer shell
 (steep density fall off, homologous velocity)
should be stronger early on, and weaken faster with time. Surprisingly, iPTF14hls shows
a very slow evolving brightness and color, broad lines (FWHM
of $\sim 10000$\,\kms), and no sign of recombination out to 600\,d.

In this letter, I show how a magnetar-powered model combined with
a standard-energy explosion of a 15\,\msun\  supergiant star can
reproduce most of the  properties of iPTF14hls. In the next section,
I discuss the numerical approach, including the treatment of the magnetar power in
the non-LTE time-dependent radiative transfer code \cmfgen\ \citep{HD12}.
In Section~\ref{sect_res}, I present the results for this magnetar-powered model,
comparing with results previously published for a standard SN II-P
(model m15mlt3; \citealt{d13_sn2p}) and a pair-instability Type II SN
(model R190NL; \citealt{d13_pisn}), and confronting with the photometric
and spectroscopic observations of iPTF14hls.
Following A17, I adopt an explosion date $MJD=56922.53$, a distance of 156\,Mpc,
a redshift of 0.0344, and I assume zero reddening.
Section~\ref{sect_conc} concludes.

\section{Numerical approach}

  The magnetar-powered SN model (named a4pm1) stems from a progenitor star
  of 15\,\msun\ initially and evolved with \mesa\ \citep{mesa3}
  at a metallicity of 10$^{-7}$. This model, which reaches core collapse
  as a blue-supergiant star, is exploded with \v1d\
  \citep{livne_93,DLW10a,DLW10b} to yield an ejecta of 13.35\,\msun,
  an explosion energy of $1.32 \times 10^{51}$\,erg, and a \nifs\ mass of 0.077\,\msun.
  Model a4pm1 has a similar He core mass and chemical stratification as model
  m15mlt3 from \citet{d13_sn2p}.
  Hydrogen dominates the ejecta composition with 7.53\,\msun.
  I adopt a strong chemical mixing (this explosion model will later be used
  for a study on SN\,1987A; Dessart et al., in preparation). Hence,
  the original low-metallicity of the envelope is erased
  by the mixing of  the metal-rich core material into the metal-poor progenitor envelope.
  At 1\,d, this model is remapped into \cmfgen\ \citep{HD12} and followed
  until 600\,d using the standard procedure \citep{d13_sn2p}.

      \begin{table}
\caption{Summary of model properties, including the progenitor
surface radius, the ejecta mass, its kinetic energy and initial \nifs\ mass,
as well as magnetar properties (for model a4pm1).
\label{tab_sum}}
\begin{center}
\begin{tabular}{l@{\hspace{2mm}}c@{\hspace{2mm}}c@{\hspace{2mm}}
c@{\hspace{2mm}}c@{\hspace{2mm}}c@{\hspace{2mm}}c@{\hspace{2mm}}
}
\hline
Model                & $R_{\star}$    & $M_{\rm ej}$ &   $E_{\rm kin}$    & $^{56}$Ni$_0$ &  $E_{\rm pm}$      & $B_{\rm pm}$  \\
                          & [\rsun]            &     [\msun]      &  [erg]                       &    [\msun]        &  [erg]                       &    [G]   \\
\hline
a4pm1     &      50          &        13.35       &        1.32(51)     &   0.077 &    4.0(50)  &  7.0(13)  \\
m15mlt3  &      501        &        12.52       &        1.34(51)     &    0.086  &    \dots  &  \dots  \\
R190NL   &    4044        &        164.1       &        33.2(51)     &   2.63   &    \dots  &  \dots  \\
\hline
\end{tabular}
\end{center}
\vspace{-0.5cm}
\end{table}

  The central feature of model a4pm1 is that starting at day one,
  I inject a magnetar power  given by
  $$
\dot{e}_{\rm pm} = (E_{\rm pm} / t_{\rm pm}) \,  / \left( 1 + t / t_{\rm pm} \right)^2    \, ,
        t_{\rm pm} = \frac{6 I_{\rm pm} c^3}{B_{\rm pm}^2  R_{\rm pm}^6  \omega_{\rm pm}^2} \,\, ,
$$
where
$E_{\rm pm}$, $B_{\rm pm}$, $R_{\rm pm}$, $I_{\rm pm}$ and $\omega_{\rm pm}$ are the initial
rotational energy, magnetic field, radius, moment of inertia, and angular velocity of the magnetar;
$c$ is the speed of light.
I use $E_{\rm pm} = 4 \times 10^{50}$\,erg, $B_{\rm pm}=7 \times 10^{13}$\,G,
$I_{\rm pm}=10^{45}$\,g\,cm$^2$ and $R_{\rm pm}=10^6$\,cm (see \citealt{KB10}
for details). This magnetar has a spin-down timescale of 478\,d.
The energy released during the first day, which is neglected, is only 0.2\% of
the total magnetar energy.
Furthermore, I assume that all the energy liberated by the magnetar goes
into ejecta internal energy (and eventually radiation) -- \cmfgen\ does not treat
dynamics.
This is a good approximation for this weakly magnetized object (see also \citealt{d18}).
In \cmfgen, I treat the magnetar power the same way as radioactive decay.
Energy is injected as 1\,keV electrons for which the degradation
spectrum is computed.
The contribution to heat and non-thermal excitation/ionization is then calculated explicitly.

   To mimic the effect of fluid instabilities\citep{chen_pm_2d_16,suzuki_pm_2d_17},
   the magnetar energy is deposited over a range of ejecta velocities. The deposition
   profile follows $\rho$ for $V <V_0$, and $\rho \exp \left(- [(V-V_0)/dV]^2 \right)$
   for $V >V_0$. Model a4pm1 uses $V_0 = 4000$\,\kms and $dV=2000$\,\kms.
   A normalization is applied so that the volume integral
   of this deposition profile yields the instantaneous magnetar power at that time.
   With this choice, the energy deposition profile influences the model luminosity
   mostly before maximum \citep{d18}.

    I compare the results to the SN II-P model m15mlt3
    \citep{d13_sn2p} and the pair-instability Type II SN model R190NL
    \citep{d13_pisn}. Model properties are given in Table~\ref{tab_sum}.

 \begin{figure}
\vspace{-0.2cm}
\includegraphics[width=\hsize]{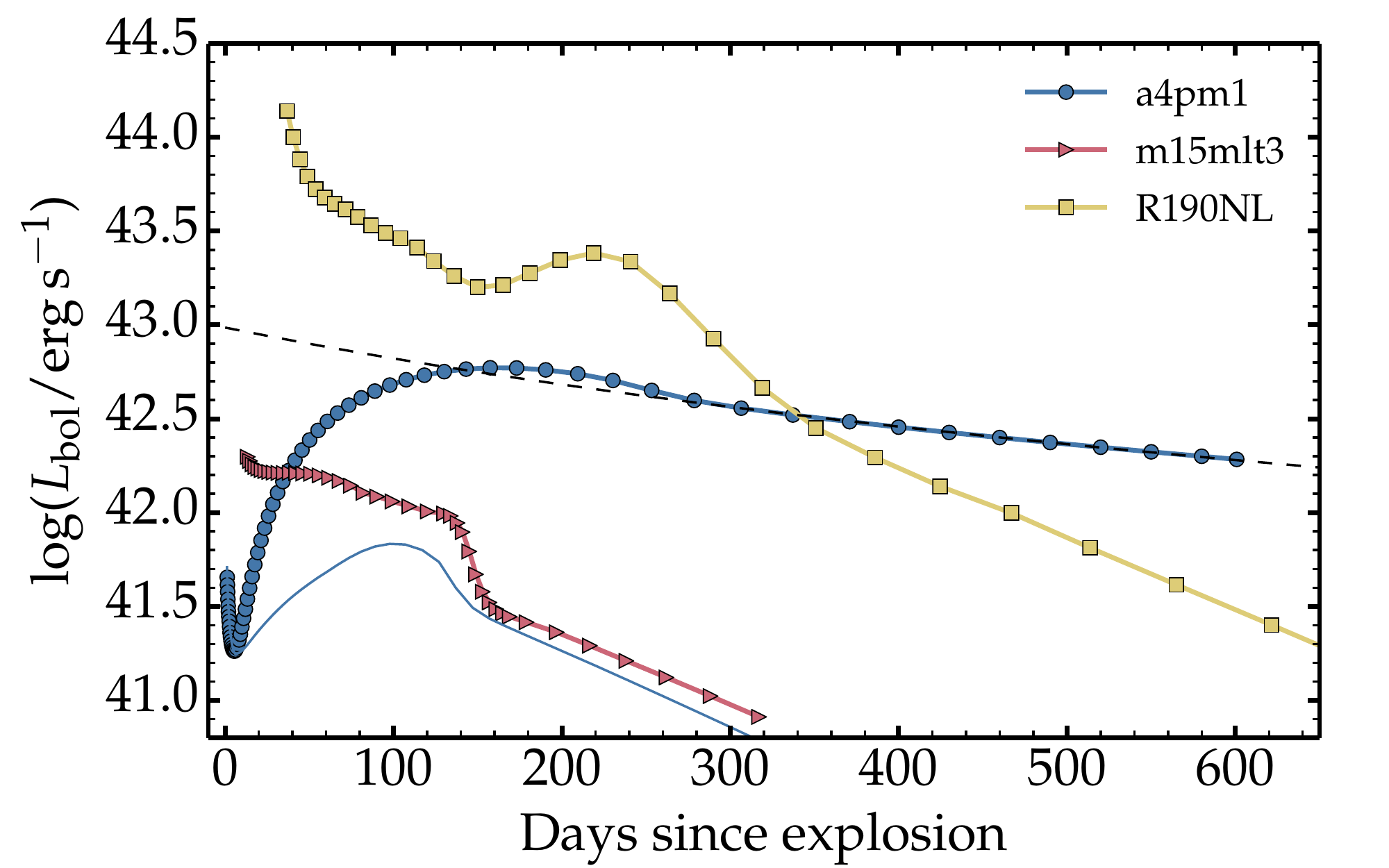}
\includegraphics[width=\hsize]{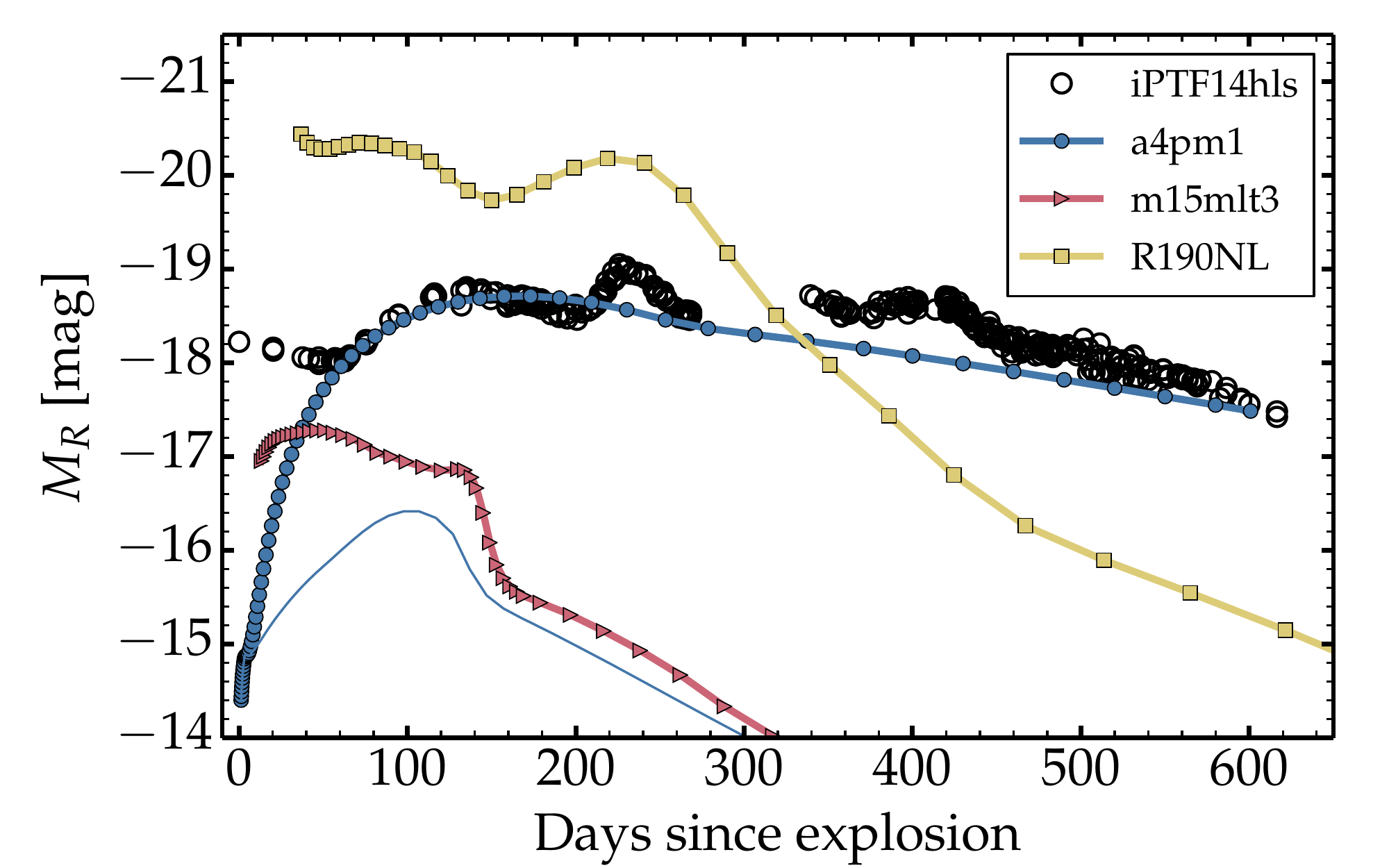}
\vspace{-0.6cm}
   \caption{Top: Bolometric light curve for models a4pm1
   (the dashed line gives the instantaneous magnetar power),
   m15mlt3 (standard SN II-P), and R190NL (pair-instability SN).
   Bottom: Same as top but showing the absolute $R$-band magnitude.
   I also add the observations of iPTF14hls corrected for distance
   and time dilation.
   The thin blue line in both panels corresponds to model a4pm1 without magnetar power.
   }
   \label{fig_photometry}
\end{figure}

 \section{Results}
 \label{sect_res}

 The top panel of Fig.~\ref{fig_photometry} shows the bolometric light curves
 for the model set.
 The magnetar powered SN is super luminous, intermediate during the first year
 between the standard SN II-P model m15mlt3 and the PISN model R190NL.
 It is the brightest of all three at late times.
 Model a4pm1 is faint early on because of the small progenitor radius.
 After $\sim 50$\,d, it follows
 closely the iPTF14hls $R$-band light curve (bottom panel of Fig.~\ref{fig_photometry}).
 The adopted magnetar power is continuous and monotonic, so it cannot explain
 the observed $R$ band fluctuations of $\sim 0.5$\,mag in iPTF14hls.
 These might indicate the intrinsic variability of the proto-magnetar. However,
 the rotation energy of $4 \times 10^{50}$\,erg and the magnetic strength
 of $7 \times 10^{13}$\,G in model a4pm1 yield a suitable match
 to the overall brightness and slow fading. The discrepancy at early times would
 be reduced by using an extended progenitor. A broader energy deposition profile
 or asymmetry might resolve this discrepancy.

\begin{figure}
\vspace{-0.1cm}
\includegraphics[width=\hsize]{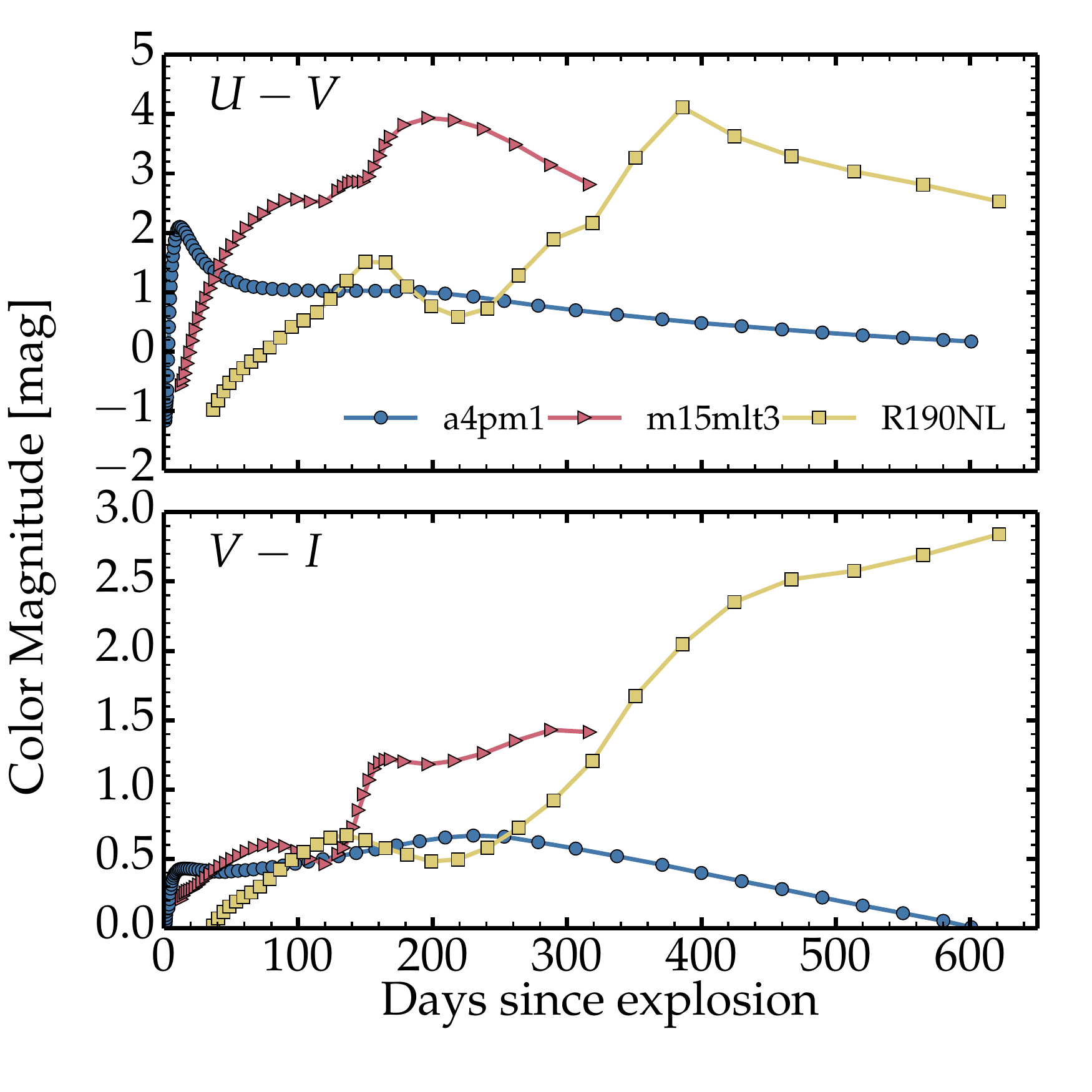}
 \vspace{-1.cm}
\caption{Same as for the top of Fig.~\ref{fig_photometry}, but now
  showing the color magnitude $U-V$  (top) and $V-I$ (bottom).
}
\label{fig_colors}
\end{figure}

 Figure~\ref{fig_colors} shows that  over the timespan $100-600$\,d
 after explosion, model a4pm1 has a weakly evolving and blue optical color,
 in contrast to the non-monotonic and strongly varying color evolution of
 models m15mlt3 and R190NL.
 Up to $\sim$\,50\,d, model a4pm1 is redder because the progenitor
 is compact rather than extended.
 This extra cooling from expansion is superseded after
 $\sim 50$\,d by the slowly decreasing magnetar power.
  Model a4pm1 follows closely the $V-I$ color of iPTF14hls, which is
  fixed at about 0.2\,mag (A17).

\begin{figure}
\includegraphics[width=\hsize]{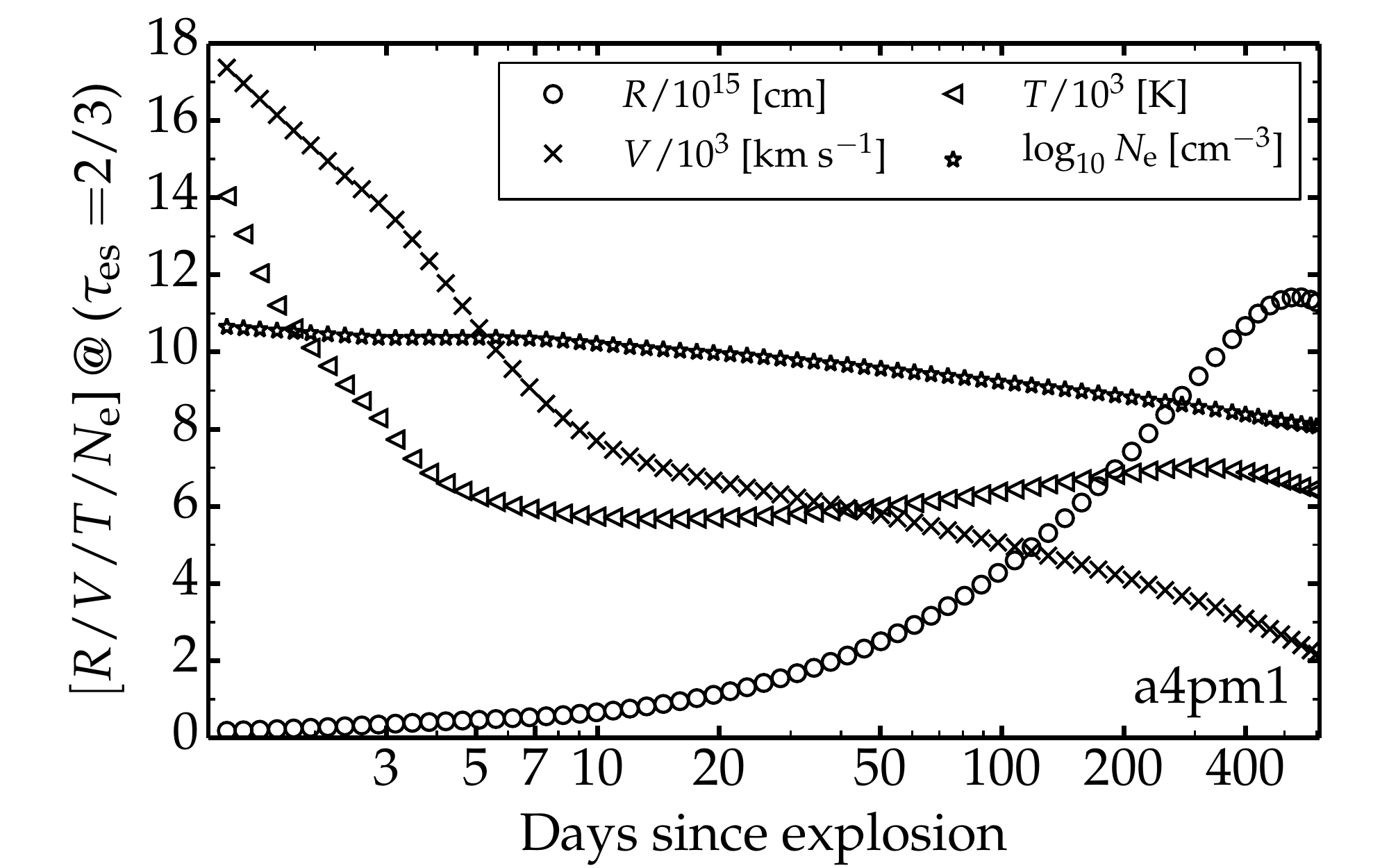}
\vspace{-0.7cm}
   \caption{Evolution of photospheric properties in model a4pm1.
The $x$-axis uses a logarithmic scale.
}
\label{fig_phot}
\end{figure}

Up to the time of maximum, this bolometric and color evolution
reflect the evolution of the ejecta properties
and of the photosphere, taken as the location where the inward-integrated
electron scattering optical depth $\tau_{\rm es}$ equals 2/3  (Fig.~\ref{fig_phot}).
In model a4pm1, the initial evolution is very rapid, as obtained
in models of blue-supergiant star explosions and inferred from the observations of SN\,1987A
\citep{DH10}. At the photosphere, the velocity (temperature)
drops from 17,300\,\kms\ (14,000\,K) at 1.2\,d down to
7,500\,\kms\ (5600\,K) at 10\,d.
After 10\,d, photospheric cooling is inhibited
and even reversed by magnetar heating and the model evolves at a near constant
photospheric  temperature of $\sim$\,7000\,K out to 600\,d.
Magnetar heating prevents the recombination of the ejecta material,
so that hydrogen remains partially ionized at all times.
This allows the photosphere of model a4pm1 to recede slowly in mass/velocity
space and to reach radii $>10^{16}$\,cm, greater than in
a standard Type II SN \citep{DH11_2p}
and comparable to model R190NL \citep{d13_pisn}.

In a standard Type II SN, the photosphere follows the layer at the interface
between neutral and ionized material (which essentially tracks the H\one\
recombination front). Recombination speeds up the recession
of the photosphere and makes the ejecta optically thin on a shorter time
scale (typically of $\sim 100$\,d) than in the case of constant ionization.
This process is mitigated by the ionization freeze-out in Type II SN ejecta
\citep{UC05,D08_time}.
In model a4pm1,  the electron scattering optical depth
$\tau_{\rm es}$ drops from $1.21 \times 10^6$ at 1.21\,d
to 1.33 at 600\,d, which is close to the value of 4.92 that would result for
constant ionization ($\tau_{\rm es} \propto 1/t^2$).  So, in model a4pm1,
the inhibition of recombination maintains the ejecta optically thick to
electron scattering for more than 600\,d. Lines of H\one\ or Ca\two\
will remain optically thick (and therefore broad) for even longer.
Between 75\% and 100\% of the magnetar power goes into heat.
Whatever remains is shared equally between excitation and ionization.
In model a4pm1, non-thermal effects are inhibited by the partial ejecta ionisation.

The photospheric evolution is not a reliable guide to
understand the SN luminosity after maximum.
The large photospheric radii combined with the large ejecta ionization
cause a flux dilution by electron scattering. The SN spectrum
may resemble a blackbody (A17), but at best diluted, with a thermalization
radius much smaller than the photospheric radius \citep{E96_epm,D05_epm}.
For example, at 250\,d,  $\tau_{\rm es}$ is 7.4, which is too small
to ensure thermalization. Instead, the conditions are nebular and the SN
luminosity equals the magnetar power (Fig.~\ref{fig_photometry}).

\begin{figure}
\vspace{-0.39cm}
\includegraphics[width=\hsize]{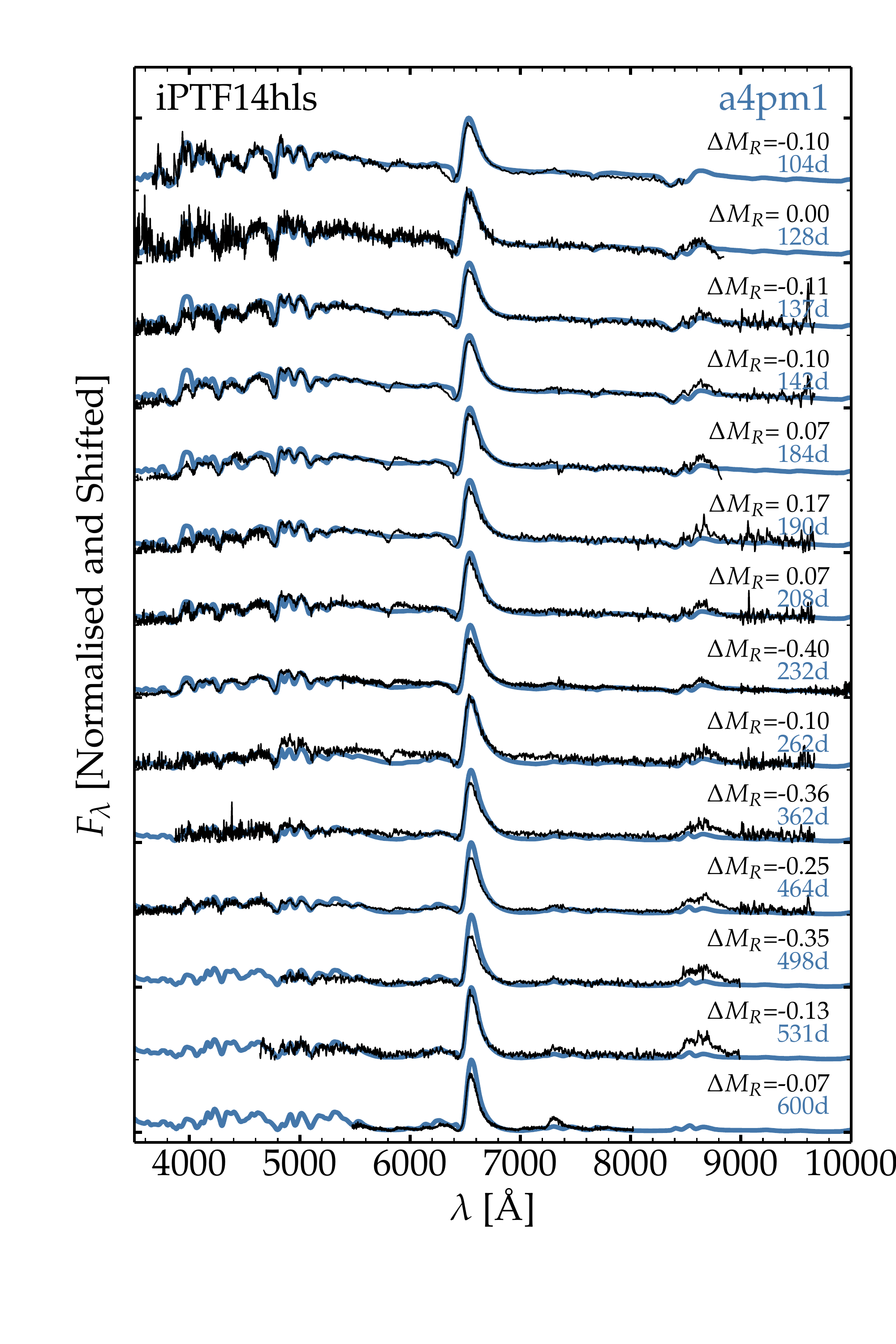}
\vspace{-1.6cm}
\caption{Comparison of the multi-epoch spectra of SLSN iPTF14hls
with model a4pm1.
Times and wavelengths are given in the rest frame.
Model and observations are renormalized at 6800\,\AA.
For each date, I give the $R$-band magnitude offset (see also
Fig.~\ref{fig_photometry}).}
\label{fig_spec}
\end{figure}

Model a4pm1 shows very little spectral evolution from 104\,d
(date of the first spectrum taken for iPTF14hls) until
600\,d (Fig.~\ref{fig_spec}), which reflects in part the fixed photospheric
conditions (velocity and temperature) after 10\,d (Fig.~\ref{fig_phot}).
The spectra show the presence of H\one\
Balmer lines, Fe\two\ lines around 5000\,\AA, the Ca\two\ triplet
around 8500\,\AA. After about 300\,d, the triplet is seen only in emission.
H$\alpha$ stays broad at all times,  and the Ca\two\ doublet 7300\,\AA\
strengthens as  the conditions in the ejecta become more nebular.
Throughout this evolution, there is little sign of the blanketing
that would appear in the optical range if the ejecta ionization dropped.
The spectral evolution of model a4pm1 is similar to that
observed for SLSN  iPTF14hls, with a few discrepancies.
The model underestimates the width of the H$\alpha$ absorption trough,
although it matches the emission width at all times.
Adopting a broader energy-deposition profile would produce broader
line absorptions (in a similar way to adopting a stronger \nifs\ mixing in Type Ibc SNe;
\citealt{d12_snibc}).

The model also underestimates the strength of the Ca\two\ emission at late times.
The feature at 5900\,\AA\ is not predicted by the model.
This is probably Na\one\,D because if it were He\one\,5875\,\AA\
one would expect a few other optical He\one\ lines, which are not seen.
Hence, our model may overestimate the ionisation. Allowing for
clumping might solve this issue \citep{jerkstrand_slsn_neb_17}.

\begin{figure}
\includegraphics[width=\hsize]{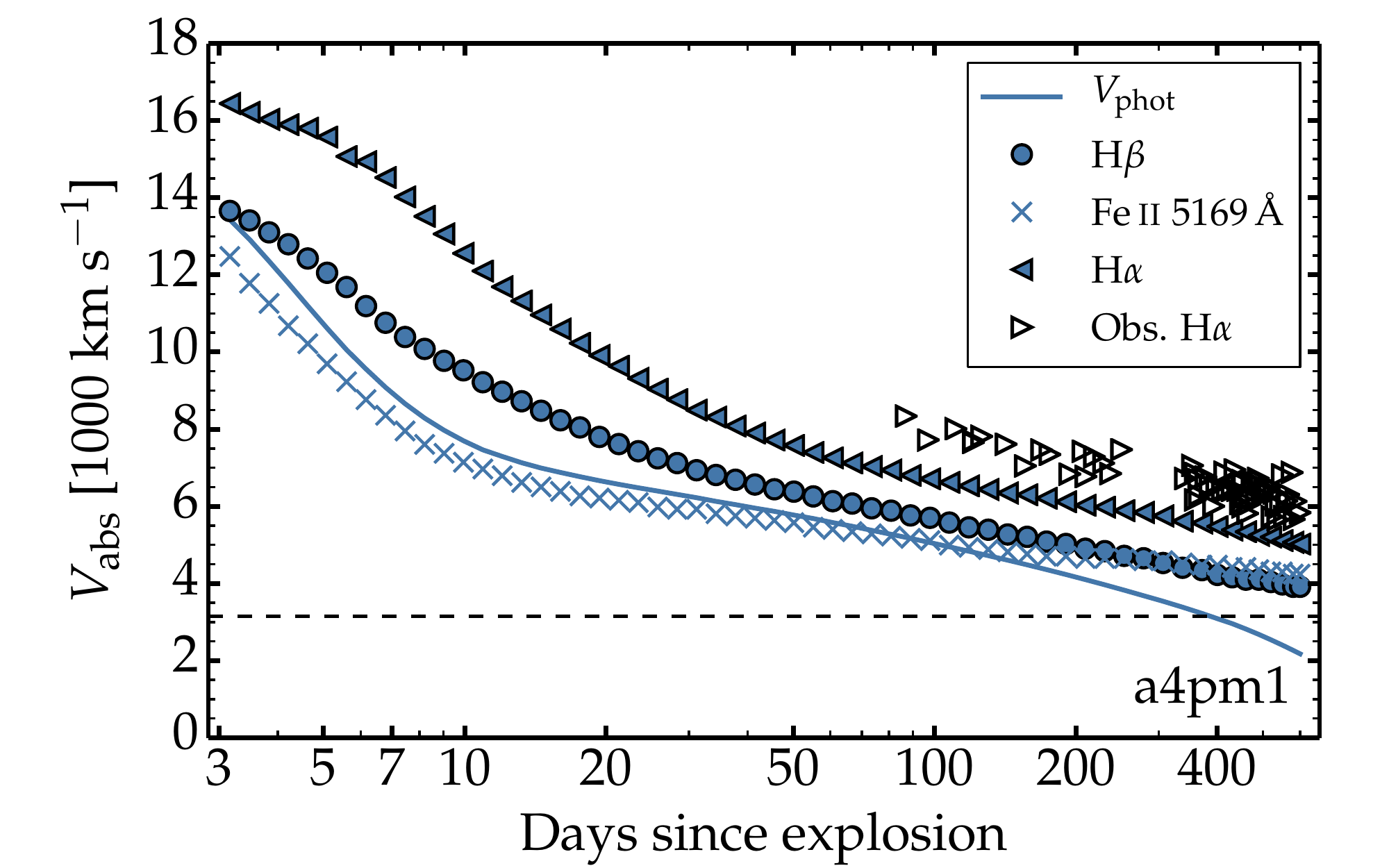}
\vspace{-0.6cm}
   \caption{Evolution of the Doppler velocity at maximum absorption
   in various lines and the photospheric velocity in model a4pm1.
   I overplot the corresponding values for H$\alpha$ in iPTF14hls.
   The $x$-axis uses a logarithmic scale.
   The horizontal line gives the ejecta velocity $\sqrt{2E_{\rm kin}/M_{\rm ej}}$.
   }
   \label{fig_vabs}
\end{figure}

The Doppler velocity at maximum absorption in H\one\ or Fe\two\ lines
is large, greater than the photospheric velocity, and does not change
much  after about 50\,d  -- the fast outer ejecta material
is scanned at early times, before the magnetar has influenced
the photosphere (Fig.~\ref{fig_vabs}).
These lines eventually form over a large volume that extends far above
the photosphere. These properties hold qualitatively even in standard Type II SNe.

\section{Conclusion}
\label{sect_conc}

In this letter,  I have presented the first non-LTE time-dependent radiative transfer
simulation of a Type II SN influenced by a magnetar.
I have shown that a magnetar-powered SN ejecta from
a $1.32 \times 10^{51}$\,erg explosion of a 15\,\msun\ supergiant star reproduces
most of the observed properties of SLSN iPTF14hls.
The modest magnetar properties ($E_{\rm pm} = 4 \times 10^{50}$\,erg,
$B_{\rm pm}=7 \times 10^{13}$\,G), combined with the standard Type II SN
ejecta properties offer an interesting alternative to the pair-unstable super-massive
star model of A17, which involves a highly-energetic and super-massive ejecta.
As discussed in \citet{d18}, a similar magnetar-powered SN, with a standard ejecta mass
and energy, may be at the origin of the SLSN OGLE-SN14-073,
for which \citet{terreran_slsn2_17} also invoke a highly-energetic and super-massive ejecta.

Hence, Type II SLSNe that show at all times a blue color, broad H\one\ spectral
lines, and a weaker-than-average blanketing,  may differ from standard Type
II SNe primarily through the influence of a magnetar.

\begin{acknowledgements}

I thank Roni Waldman for providing the progenitor used for model a4pm1.
This work utilized computing resources of the mesocentre SIGAMM,
hosted by the Observatoire de la C\^ote d'Azur, France.
This research was supported by the Munich Institute for Astro-
and Particle Physics (MIAPP) of the DFG cluster of excellence
``Origin and Structure of the Universe".

\end{acknowledgements}

\end{document}